\documentclass[a4paper,12pt,superscriptaddress]{article}
\usepackage[english]{babel}
\usepackage{amssymb}
\usepackage{amsfonts}
\usepackage{amsmath}

\usepackage{graphicx}
\usepackage{subfigure}

\usepackage[round,comma,authoryear]{natbib}

\begin{document}

\title{Models and Simulations in Material Science:\\Two Cases Without Error Bars}
\author{Sylvia Wenmackers\thanks{Theoretical Philosophy Group, Faculty of Philosophy, University of Groningen, The Netherlands. E-mail: s.wenmackers@rug.nl}~\ and Danny E.~P.~Vanpoucke\thanks{Department of Inorganic and Physical Chemistry, Faculty of Science, University of Ghent, Belgium. E-mail: danny.vanpoucke@ugent.be}}
\date{}
\maketitle

\begin{abstract}
We discuss two research projects in material science in which the results cannot be stated with an estimation of the error: a spectroscopic ellipsometry study aimed at determining the orientation of DNA molecules on diamond and a scanning tunneling microscopy study of platinum-induced nanowires on germanium. To investigate the reliability of the results, we apply ideas from the philosophy of models in science. Even if the studies had reported an error value, the trustworthiness of the result would not depend on that value alone.
\end{abstract}
%
%
\section{Introduction}\label{sec:Intro}
Material science is an interdisciplinary branch of science, mainly populated by experimental and computational physicists, chemists, biomedical scientists, and engineers. They work in large research groups and consortia, on topics such as semiconductors, nanostructured materials, photovoltaic cells, and bioelectronics. When philosophers discuss physics, they often focus on issues in theoretical physics, rather than topics in solid state physics and material science. Yet, this type of research brings about its own philosophical questions, some of which we will address in this paper. Like all empirical sciences, material research requires the confrontation and combination of experimental findings with theoretical entities, including models. Other issues, however, are strongly case-specific. For that reason, we structure our paper around two case studies. Only recently did philosophers shift their focus from scientific theories to models, and from pure, theoretical branches of science to experimental and interdisciplinary fields. The topic of the current paper---models in material physics---remains wide open for exploration.

Statisticians typically deal with `empirical models': mathematical models that describe data sets, without a scientific background theory about how the data relate to reality. Experimental physicists, on the other hand, usually analyze their data on the basis of `physical models', which are derived---in a non-trivial way---from physical theories, such as classical mechanics, quantum mechanics, optics, or the like. In material science, statistical and physical modeling go hand-in-hand. Therefore, in the context of this article, it does not suffice to discuss empirical models alone. We need a broader framework that encompasses all types of `models  in science', including empirical as well as scientific models.

In section~\ref{sec:PhilOfModels}, we review existing work on models in science by philosophers. Philosophy of science does not claim to offer a general classification of models, so it should not come as a surprise when we encounter some lacunae along the way.

In section~\ref{sec:TwoCases}, we deal with two cases in material science which were the topics of the doctoral study of the first and second author \citep{Wenmackers:2008,Vanpoucke:2009}. For each case, we introduce the research question, the methodology, and the obtained result. In both cases, the result cannot be stated with an estimation of the error (albeit for different reasons), leading to questions about the reliability of the results. To answer these questions, we look into the particulars of each case and analyze the work flow. We observe that they involve a complex interplay of scientific theories and models. By making explicit which models are involved, we try to establish the trustworthiness of the results of both cases.

In section~\ref{sec:Outro}, we give a general discussion of the cases and distil common conclusions, most of which hold for material science in general.


\section{Philosophy of models in science}\label{sec:PhilOfModels}
Recent case studies on the use of models in specific branches of science---such as \citet{Guillemot:2010} and \citet{Parker:2010} on climate science, \citet{BevenWesterberg:2011} on hydrology, and \citet{GruneYanoff:2009} and \citet{PhanVarenne:2010} on sociology---show the diversity of models, which involve creative and highly case-specific elements. Hence, any classification of models is necessarily incomplete: the distinctions in it are not mutually exclusive, and the way of categorizing never unique. This wild variation, which is intrinsic to the topic of modeling, may be part of the reason why philosophers of science have traditionally focused on scientific theories rather than models. Any philosophy that would claim to offer an all-encompassing theory of models in science should be treated with proper suspicion. Moreover, it would be a pity if such a classification would be misunderstood as a normative device, thus becoming a self-fulfilling prophecy, blocking some of the intellectual plasticity badly needed in the practice of modeling.
\subsection{Semantics, epistemology, and ontology of models}
Although philosophy of science does not offer a full-blown theory of models in science, it does provide us with some tools for the conceptual analysis of models that are used in statistics (such as mathematical models and models of data) and in material physics (such as material models). In their encyclopedia article, \citet{FriggHartmann:2006} discuss the scientific models from the perspective of philosophy of science. Here, we give a short overview of their take on the semantics, epistemology, and ontology of models in science, to enable easy reference in the discussion of the case studies.

The semantics of scientific models deals with the question: \emph{what do models represent?} One way of answering this question, is to distinguish between models that (a) represent a part of the world, and (b) those that describe a part of a theory. This is a useful distinction, even though some models may do (a) and (b) simultaneously. In case (a), the relevant part of the world is called the `target system', which may consist of (a1) phenomena or (a2) data (again, not mutually exclusive). In case (a1), the corresponding model is called a `model of phenomena', in case (a2), a `model of data'. Models of data are the most relevant category of models in statistics. Observe that scientists almost never apply statistics to `raw data': there always is an element of selection, and sometimes correction, involved. In other words, the data sets that are used in a statistical analysis are `clean data', which serve as a model of the raw data. It is also in the context of (a2) that the curve-fitting problem appears.

The epistemology of scientific models deals with the question: \emph{how can we learn from models?} It seems clear that we can learn how a model works: by working with a simple mathematical model for mechanics, for instance, we can learn how to solve the equations of motion, thereby learning something about the model of motion. But ultimately, we want to learn something about the world; the model is only a means to that end.\footnote{There are exceptions. Although almost all models are intended to represent something in reality, some mathematical models are simplified so much that they clearly do not represent anything in reality. These `toy models' are studied for their own sake, rather than to learn about the world.} Given that a model is distinct from its target system, it is not immediately clear how we can learn about the world by manipulating the model. \citet{Hughes:1997} proposes a way to describe the process of learning about the world via a model. He distinguishes between three phases in the process: denotation, demonstration, and interpretation. (In practice, these steps need not be clearly distinct and are not to be understood as three consecutive steps: going back-and-forth between the steps may occur.) In the denotation phase, certain features of the world are related to elements of the model. For example, it is decided that the symbol $v$ in the model shall represent the velocity of an object in the world. The demonstration phase takes place entirely in the model, and consists of activities such as learning, predicting, and explaining. In the interpretation phase, what has been learned (predicted, explained, \textit{etc.}) in the model is translated into corresponding statement about the world (by means of the relations established in the denotation phase). So, in a sense, this is the opposite move of the denotation phase.

The ontological question concerning scientific models is: \emph{what are models?} \citet{FriggHartmann:2006} consider a distinction between material models and ideal models. Examples of material models, also called iconic models, include scaled-up or scaled-down replicas of the target system, and samples (or specimens). A sample is a representative element of a larger group of elements of interest. One may think of biological or medical studies, in which in order to study an entire species or a certain type of patients, respectively, one or a limited number of representative elements (members of the species or patients) is studied. The sample is also important in material physics, as we will see in the case studies: there, `sample' has a slightly different meaning, as it is something that is specially prepared, rather than selected. Ideal models are immaterial models, conceptual by nature, such as descriptions, thought experiments, and mathematical models. The same ideal model may be represented in multiple ways: as a text, a schematic drawing, a set of equations or matrices, and so on. An important subset of ideal models in statistics, material physics, and all other branches of science are mathematical models. In this paper, by `simulation' we shall refer to an instance of a mathematical model.

\subsection{All models are wrong}
After this short review of the questions considered by \citet{FriggHartmann:2006}, we already have a good basis to reflect about the question implied by the conference title: \emph{Are all models wrong?}
As a first observation, notice that only a number of features of the target system are considered to be relevant for a given research context and that the model only represents those. Hence, all models are approximations. If neglecting all other features of the target and its relations to other parts of the world turns out to be inconsequential to the features under study, the model is useful. No model is exactly like its target in every aspect; so, under the highest epistemic standards, all models are wrong.

The intended function of a model is to create something that is easier to understand or to manipulate than the target. Many additional approximations are built into the model intentionally: additional steps away from reality may be required to make the model better suited for educational or computational purposes. Idealizations are a special type of approximations: parameters that are known to have some influence on the target system may still be left out of the model because otherwise the model becomes too complex to be useful.
Whereas increasing the difference between a model and its target system may have the advantage that the \emph{model} becomes easier to study, studying a model is ultimately aimed at learning something about the \emph{target system}. Therefore, additional approximations come with the cost of making the correspondence between model and target system less straightforward. Ultimately, this makes the interpretation of results on the model in terms of the target system more problematic. We should keep in mind the advice of \citet[p.~163]{Whitehead:1920}: ``Seek simplicity and distrust it.''

A `good model' is to be understood as a model that achieves an equilibrium between being useful and not being too wrong. The usefulness of a model is clearly context-dependent; it may involve a combination of desired features such as being understandable (for students, researchers, or others), achieving computational tractability, and other criteria. `Not being too wrong' is to be understood as `not being too different from reality'. However, we can never establish the exact difference between the target system and our model for it: we can measure differences between models (\textit{e.g.}\ compare how well they match a set of data), but we can never be sure that the parameters we have selected to do so apply to reality at all. It is well-know by statisticians that there are different measures for the distance between models (such as the likelihood) and that different information criteria may rank models differently.
One may try to summarize the overall quality of each model by a single figure of merit, by attaching relative weights to the various criteria. The choice of these weights, however, is again open for debate, and even if a consensus could be reached on this matter, \textit{ex aequo}'s are possible.


\section{Two cases in material science}\label{sec:TwoCases}

\subsection{Case A: Ellipsometry of DNA on diamond}\label{sec:CaseElli}
Our first case study in material science is based on the paper of \citet{Wenmackers_etal:2008}; it is part of a larger project which focuses on diamond coatings as a platform material for biosensors, in particular DNA sensors \citep{Wenmackers_etal:2009}. Single stranded (ss-) DNA probes with a known sequence are covalently attached to the sensor surface with one end. When patient DNA is brought into contact with the bio-functionalized surface, the patient DNA will only bind to the probes, thereby forming double stranded (ds-) DNA, provided that they are of the complementary sequence. When binding is detected, the sequence of the patient DNA can be deduced from it.

An important parameter that influences sensor performance is the orientation of the DNA probes: only if the probes are directed upwards, extending from the surface into the buffer fluid, they are available for hybridization to patient DNA. In this case study, the main research question is: ``What is the average angle between the backbone of the probe DNA and the plane of the sensor surface?''

The characterization method chosen to answer this question was spectroscopic ellipsometry (SE). Ellipsometry monitors the change in polarization when light interacts with a material interface and is highly surface sensitive. In an ellipsometry measurement, two parameters are recorded: the amplitude ratio, $\tan(\psi)$, and the phase shift, $\Delta$ \citep{Arwin:2000}. Using light of a fixed wavelength, ellipsometry can be used to determine a single unknown layer thickness. By performing the measurement for a whole spectrum of wavelengths, one can derive additional information from the data, such as the orientation of the molecules in a layer.

Various structures in DNA molecules absorb at specific wavelengths in the ultraviolet to visible (UV-Vis) range. In particular, DNA can absorb wavelengths of 261~nm (corresponding to an energy of 4.74~eV); this absorption is associated to a $\pi$-$\pi^*$ electronic transition (in which the orbitals change from bonding to non-bonding) in the bases of the DNA. The conformational sensitivity of the technique originates from the spatial orientation of the dipole moment associated with the selected excitation.

SE is an indirect measurement technique. In the words of \citet[p.~2--3]{Fujiwara:2007}: ``[e]llipsometry data analysis requires an optical model defined by the optical constants and layer thicknesses of a sample. In an extreme case, one has to construct an optical model even when the sample structure is not clear at all.'' An important optical constant is the complex refractive index (for a specific material and for a fixed wavelength) $N = n + ik$, with $n$ the real-valued refractive index and $k$ the extinction coefficient.

\begin{figure}[!tb]
\centering
\includegraphics[width=8cm]{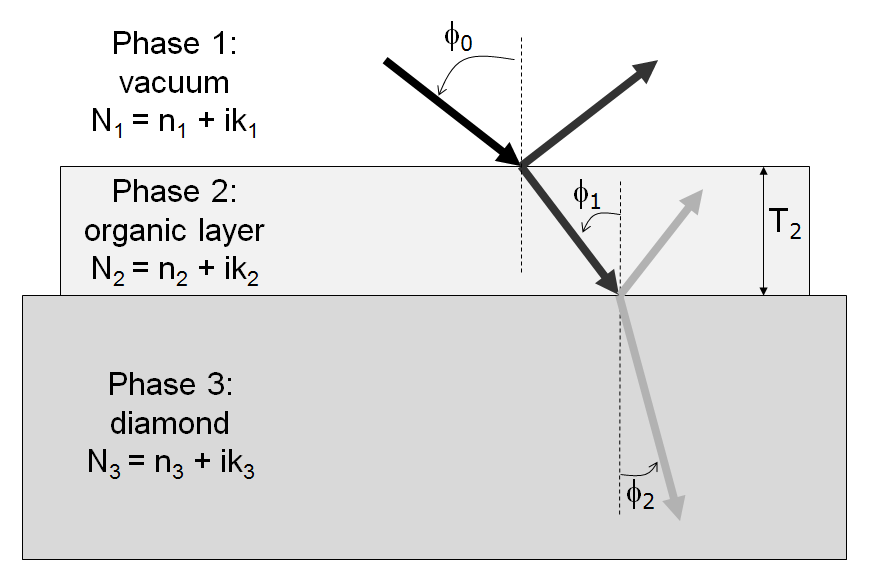}
\caption{Three-layer optical model for an organic layer on diamond.}\label{Fig:OpticalModel}
\end{figure}

In this case, the model consists of three layers, as depicted in Figure~\ref{Fig:OpticalModel}: the diamond phase (bulk), the organic layer, and the ambient phase (vacuum). By comparing the in-plane and out-of-plane contributions to the 261~nm absorption of the DNA bases and based on the optical model, the average angle of the DNA bases can be calculated. The complement of that angle, $\alpha$, is taken to be the average angle of the DNA molecules, the sought parameter. The result is that $\alpha = 45^\circ$ for ss-DNA of 8 bases long, $\alpha = 49^\circ$ for ss-DNA of 36 bases long, and $\alpha = 52^\circ$ for ds-DNA of 29 base pairs long.

In SE studies, the goodness-of-fit of the spectra of $\tan(\psi)$ and $\Delta$ as a function of the wavelength is usually assessed both quantitatively and qualitatively. For the quantitative part, fitting errors are calculated based on a least-squares method \citep[p.~199]{Fujiwara:2007}. Additionally, a qualitative assessment is done by plotting the experimental spectra together with the simulated spectra on one graph, to check whether they have the same overall shape, in particular in the region with the absorption(s) of interest. Because of the complexity of the calculations, however, ellipsometry studies typically do \emph{not} report an error for the obtained values of the parameters in the optical model and the derived parameters, such as the tilt angle, $\alpha$.

This prompts the question: how reliable is the result? Is the probe ss-DNA of 8 bases \emph{really} oriented under $45^\circ$ as compared to the plane of the sensor surface? Although $\alpha$ is a parameter derived from the optical model, it is interpreted in a straightforward and realistic way: as the average angle of the length-direction of the DNA molecules on the diamond surface. In order to investigate whether this direct identification is warranted or not, we have to inspect all the measurements, models, and assumptions that lead to the determination of this parameter; this is done in section~\ref{sec:DiscAB}.


\subsection{Case B: STM of nanowires on germanium}\label{sec:CaseSTM}
Our second case study is based on the paper of \citet{VanpouckeBrocks:2010a}. It is part of a larger research project which aims at the miniaturization of computer chips: atomically small wires and devices are the ultimate goal in this endeavor. The study focuses on the mono-crystalline germanium surface Ge(001), and its behavior when small amounts of platinum (Pt) are deposited. Scanning tunneling microscopy (STM) observations show that after high temperature annealing of such samples, single-atom wide nanowires (NWs) appear on the surface \citep{Gurlu_etal:2003}.

STM is a type of scanning probe microscopy in which an atomically sharp tip is brought in close vicinity of the studied surface and then scans over it. A bias applied between the tip and the surface results in a tunneling current. The tunneling current is kept constant by a feedback loop which adjusts the height of the probe, resulting in a two-dimensional height map. The height map can be interpreted as a direct reflection of the topology of the sample surface.
Although STM can determine the surface topography with atomic resolution, it cannot determine which chemical elements are present. Due to this lack of chemical sensitivity, the experimenters have to rely on their intuition and circumstantial evidence \citep[such as adsorption experiments with CO, ][]{Oncel_etal:2006} to speculate about the chemical nature of the NWs. Although they refer to the structures as `platinum nanowires', it would be safer to call them platinum-induced nanowires until further information on their composition is available.

\begin{figure}[!tbp]
\centering
\includegraphics[width=12.5cm]{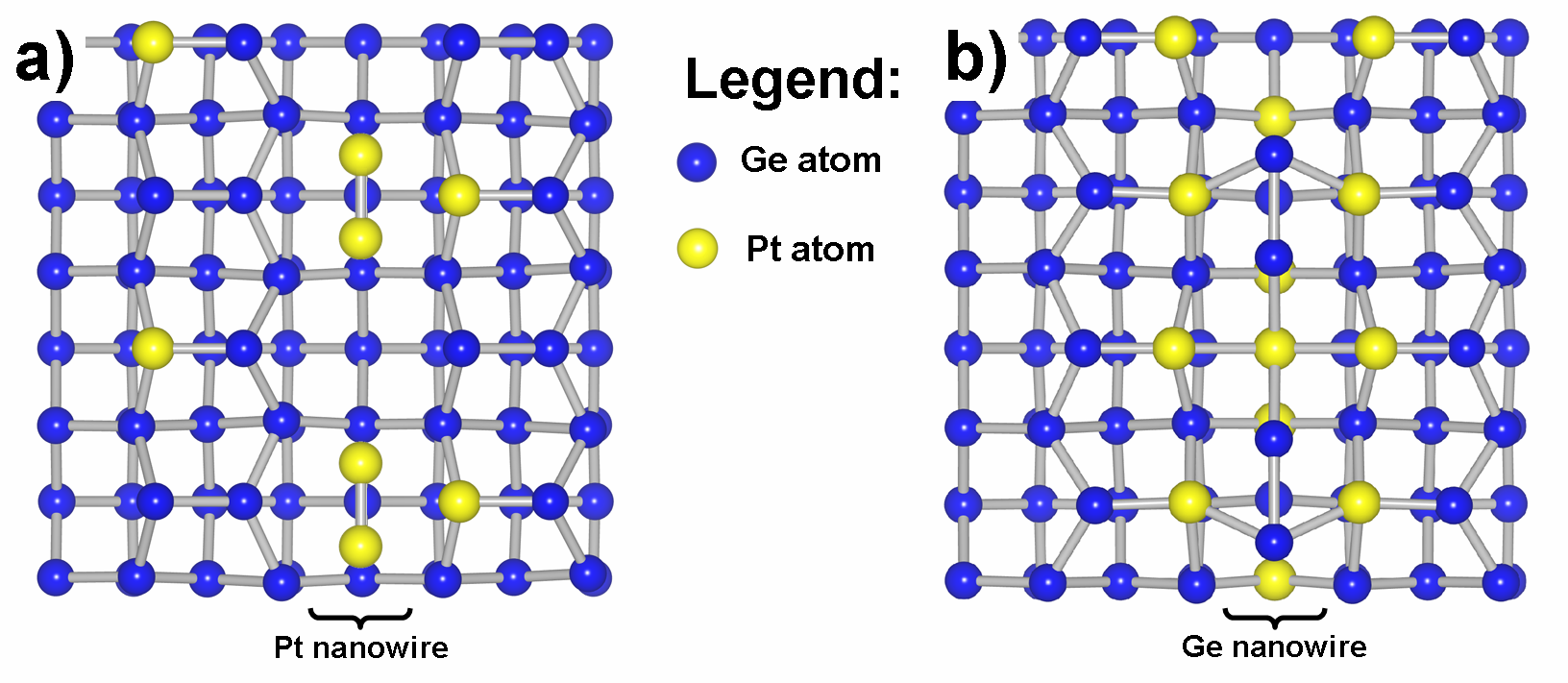}
\caption{Atomistic models of Pt-induced nanowires on Ge(001). Ge/Pt atoms are indicated in blue/yellow. (a) The experimental model involves Pt dimers on top of the Ge surface. (b) In the final theoretical model, the Pt atoms diffuse into the subsurface, pushing up some Ge atoms. These dislocated Ge atoms appear as NWs in the simulated STM images.}\label{Fig:AtomModel}
\end{figure}

The aim of the theoretical study of this system is to determine the chemical nature of the NWs and to clarify the mechanism of their formation. In order to do this, a set of atomic structures is proposed; one such atomic structure is shown in Figure~\ref{Fig:AtomModel}.a. Starting from each structure, a theoretical STM image is calculated and compared to the experimentally obtained images. The features that match the experimental images best are then used to create a new set of atomic structures, for which STM images can be calculated. This loop is iterated a few times, leading to the model structure of Figure~\ref{Fig:AtomModel}.b: the simulated STM images based on this model show all the relevant features present in the experimental images. The conclusion of this study is that the deposited Pt atoms exchange positions with Ge atoms in the surface layer. This results in Ge NWs on a Pt-modified surface. In conclusion, the observed nanowires consist of germanium.

In this case, the research question and the corresponding answer are of a qualitative rather than quantitative nature. Although the analysis does rely heavily on quantitative information, no quantitative error can be given. Again, the question is: how reliable is this result? Are the observed nanostructures \emph{really} made of germanium?


\subsection{Scientific models in cases A and B}\label{sec:DiscAB}
As we have seen in section~\ref{sec:PhilOfModels}, a model can describe a target system, which is to be understood as a part of the world. The targets of these case studies are (data associated with) a diamond-based DNA sensor and a nanowire-based chip: objects which do not actually exist (yet) and therefore are not parts of the world. This situation, in which the target system itself is an ideal model, is quite common in material science and engineering.

What does exist in the material world are samples, which serve as precursors or prototypes of the DNA sensor or NW-chip in development. Should we consider these samples as the target system, then? No, the sample in turn is a material model, representing something more than itself: it is thought of as a typical representative of a larger class of samples that are or could be produced by the same protocol.

The protocol is a list of manipulations that have to be performed in order to produce a sample (diamond with probe DNA or Ge with NWs). Although it is likely that the protocol itself requires further optimization, it does serve as an ideal model: it is a model of how the actual preparation, which results in the sample that is actually measured, should be carried out.

For the unaided eye, it is usually not possible to verify if the protocol has worked. In case A, the SE measurement required scarce beam time at a synchrotron facility (BESSY II in Berlin). Therefore, preliminary tests were performed to verify the presence of DNA on the samples: fluorescence microscopy, atomic force microscopy (AFM), and contact angle measurements \citep{Wenmackers_etal:2008}. To avoid contaminating the actual samples, it is possible to prepare additional samples in parallel with the ones that are to be used in the actual SE experiment. The additional control samples serve as a material model for the actual sample, which in turn is a model for the DNA sensor, which is an ideal model of a device that does not exist yet. The disadvantage of this option is that one cannot verify the inter-sample variability and has to assume that the samples are essentially equal; in terms of statistics, the various samples should at least be describable by the same distribution. Notice that doing the preliminary tests on the actual sample does not always get rid of the representativity problem: most microscopic techniques can only image a small part of the total sample surface, and thus it has to be assumed that the tested area was representative for the whole sample.

Once the samples are prepared and the initial tests have proven positive, the actual characterization method can be applied. Both UV-Vis SE and STM require a complex apparatus. Although the actual measurement process is an interesting topic in its own right, we just assume here that this part is completed successfully: one has collected many spectra or images---the raw data---typically more than one will be able to analyze. The researcher now has to select the `good' spectra or images, according to criteria which are qualitative rather than quantitative in nature. The selected spectra or images are called the clean data, and can be regarded as a model of the raw data.

We have seen that case A relies on an optical model, consisting of three-layers, and that case B involves an atomistic model: a specification of the positions of the various atoms and their chemical species. Let us now examine how these models are constructed. They are clearly not empirical models, since they are not constructed solely by working with the clean data. Rather, the models are constructed largely independent of the data and depend on an interplay between other sources of information: (1) on the experimental method and (2) on the sample. Type (1) information tells us which is the relevant background theory (optics for case A, quantum mechanics for case B) and how the measurement apparatus is designed. Type (2) information includes knowledge of the protocol, logbook entries from the actual execution of the protocol, and the outcome of preliminary tests. From these different sources of information, the models are derived. It is clear that there is no algorithm for turning the information into a suitable model. There are degrees of freedom, and within the bounds of this freedom, experience and heuristics play a role in the development of the model.

For case A, the optical model can be considered to be a mathematical model with parameters. Whereas the number of variables in empirical models of statistics is usually determined by employing an information criterion (IC), this practice is not applied in physics. The background theory leaves room for a number of parameters, which all have a physical interpretation, such as a layer thickness or the spatial orientation of electrical dipole moment associated with a specific absorption. It may be the case that the number of parameters is too large to be calculated from a single spectrum. In that case, additional measurements may be done to gather more information. (For example, measure a bare diamond surface to establish the optical constants for the bulk material.) If the number of variables is still too large, it can be reduced further by idealizing the model. In the optical model of case A, the layer roughnesses of diamond and DNA are deliberately excluded: the model is made less realistic in order to make the calculations lighter (or to make them possible at all). In other words, idealizations and other approximations make the model more wrong, in order to make it (more) useful. However, such deliberate `errors of omission' \citep [\textit{cf.}][]{BevenWesterberg:2011} may backfire in the interpretation step, as we will see in section~\ref{sec:realismAB}.

Underlying the atomistic model for case B there is a realistic assumption: the assumption that there exists a certain spatial organization of atoms which is the cause for the observed structures in the STM images. Based on the experimental conditions, it is also assumed that the atoms belong only to two different species (Ge and Pt).
Although the model depends on the Pt/Ge-fraction, this numerical variable does not fully determine the model. Proposing an atomic model for a NW is not a matter of changing the composition of a fixed matrix, since the available positions (in the crystal and on surface adsorption sites) change with the composition. Hence, the model in case B is not parametric in the usual sense.
The simulated STM image is calculated based on quantum mechanics (QM), a powerful and well-established physical theory. Yet, the practical application of QM requires many idealizations and approximations to be made. Firstly, density functional theory (DFT) is an approximative theory used in order to make QM practically applicable to many-body systems (such as the Ge surface of case B).
Secondly, instead of explicitly taking tunneling into account, the simulation only calculates an isosurface of charge density, which is assumed to be equivalent to the height profile in a constant-current STM image. This is equivalent to modeling the STM tip as an ideal point source, but with the advantage of requiring less computation.
A third way of reducing the computational cost is using only a small number of atoms and applying periodic boundary conditions. The resulting surface then consists of exact copies of the same small surface unit, clearly an unrealistic model assumption.

The models can be represented in different ways. In case A, the optical model can be represented as a schematic drawing (as in Figure~\ref{Fig:OpticalModel}), or as $4 \times 4$ Mueller matrices, for instance. The model of case B can be represented as a list of coordinates, to be used by a computer algorithm, or as a ball-and-stick image (as in Figure~\ref{Fig:AtomModel}). We regard these different representations as the same model. The use of these models for simulation and fitting is discussed in the two following subsections.


\subsection{Statistical procedures in cases A and B}\label{sec:StatAB}
By assuming certain values for the model parameters, the optical model can be used to obtain simulated spectra. The fitting procedure consists of identifying those parameter values that fit the experimental spectrum (from the clean data) best. Additional calculations on these best-fit parameter values ultimately lead to a value for $\alpha$. Stripping away the particulars, leads us to this abstract presentation of case A:

\begin{description}
  \item[(A1)] Assume a model $M$ with the main parameter of interest, $\alpha$.
  \item[(A2)] Given the protocol $P$ that generates the data and given the model $M$, estimate $\alpha$ by $\hat{\alpha}$.
\end{description}

Since the model $M$ may be wrong, the uncertainty about $\alpha$ is larger than the internal error in $\hat{\alpha}$. The latter observation happens on a meta-level as compared to the statistical modeling of steps (A1) and (A2); typically, on such a meta-level less quantitative tools are available and we have to resort to qualitative modes of critical thinking.


For case B, the algorithm is more complex:

\begin{description}
  \item[(B1)] Experimental data, $D$, are collected under (uncertain) experimental conditions, $\theta$. Initialize trial counter: $i=1$.
  \item[(B2)] Make assumptions, $\theta'_i$, about the experimental conditions $\theta$ and propose a model $M_i$ that is consistent with these assumptions. At $i=1$, only the partial information on the experimental conditions is used to generate ($\theta'_i$,$M_i$); for $i>1$, previous outcomes of step (B4) are also taken into account.
  \item[(B3)] Use $M_i$ to simulate data, $D'_i$.
  \item[(B4)] Compare features of $D'_i$ to features in the experimental data $D$.
  \item[(B5)] If the match in (B4) is insufficient, propose changes to $M_i$ for the next trial, add 1 to the trial counter, $i$, and return to step (B2). If the match is sufficient, accept $M_i$.
\end{description}

Although we assume that there exists a true model (see next subsection), there is no guarantee that it is the selected model $M_i$. The reason is twofold: we start from incomplete information about $\theta$ and we can only explore a small fraction of all possible models. Yet, we do not know how to quantify the likelihood of $M_i$.

\citet{Marjoram_etal:2003} discuss the situation in which no likelihoods are available. In particular, algorithm F with step (F3) replaced by (F''3) in their paper resembles the above structure of our case B. Let us briefly discuss three key elements.

Firstly, the generation of a simulated STM image in step (B3) corresponds directly to step (F2).

Secondly, the experimental and simulated data in case B consist of STM images (height maps), but step (B4) mentions `features' of these experimental and simulated data. Examples of such features are the relative positions of peaks and valleys, their relative heights, and the presence of symmetric bulges. Experience learns us that small changes in the modeling settings result in few changes in these features of the resulting STM images, although the absolute height and shape of the peaks may change severely. A similar idea is known in the statistics literature as Approximate Bayesian Computation (ABC): the relevant features are called a `set of statistics'. The idea of ABC, also present in step (F3''), is to measure the distance between the statistics of the experimental and simulated data, $S$ and $S'$, rather than between the data, $D$ and $D'$, directly.
Whereas (F3'') assumes the existence of a metric to compare the difference between the experimental and simulated data, in our case, the resemblance was judged by a human observer, resulting in a qualitative rather than a quantitative judgement.
It is straightforward to compute the distance between two height maps (an experimental and a simulated STM image), but a quantitative comparison of the relevant features in two such images is currently out of reach; it would require substantial input from the domain of computer vision.

Thirdly, step (B1) and (B5) embed steps (B2)--(B4) in a cyclical learning loop. Likewise, step (F1) and (F4---(F5) embed steps (F2) and (F''3) in a Markov chain Monte Carlo (MCMC) method. Unlike in MCMC, in our case the new model has to be constructed by hand, something which could be considered as a poor man's MCMC method. The reason for this is twofold: the phase space is huge and our models are not parametric (as discussed in section~\ref{sec:DiscAB}). Also observe that $\theta'_i$ are assumptions about the experimental conditions, not just model parameters in the usual sense: they leave the associated model, $M_i$, underdetermined. Since case B is not based on a parametric model, it also does not allow us to express the probability of the selected model in terms of a prior over model parameters as in step (F4).

In conclusion, although the structure of our case B resembles the F''-scheme as discussed by \citet{Marjoram_etal:2003}, it lacks the quantifiable elements (algorithmic transition kernel in the MCMC, distance metric between the statistics, prior over model parameters). However, future developments, in computer learning and automatized image recognition, could supplement our current method with quantification.
If the genetic algorithm for automating the procedure existed, it could associate a figure of merit to the selected atomistic model. But even then, a broader discussion of the reliability of the result would be necessary. For instance, it would still depend on our implementation to which features the computer is sensitive, and how it weighs their relative importance. Although it is a positive step to objectify the criteria, the criteria for choosing a particular set of weighted criteria remain open for discussion. This shows that automating model selection processes and quantifying their result by figures of merit (or error) does not eradicate all the associated philosophical questions; they merely obfuscate them.


\subsection{Physical realism in cases A and B}\label{sec:realismAB}
Let us now return to the question regarding the physical interpretation of the statistical results. Is the estimate for the indirect parameter $\alpha$ in case A \emph{really} the tilt angle of the DNA molecules? Does the selected atomic model in case B correspond to the \emph{real} configuration of atoms in the NW?

Given the complexity of the whole process (from sample preparation, over measurement, to data analysis), there are many occasions for something to go wrong \citep[\textit{cf.}][]{Ioannidis:2005}. Reliability of the results requires at the very least that the experimental observations are reproducible. This minimal robustness condition is satisfied for both case A and case B. In case B, also the theoretical model was tested further by verifying that its results are consistent with the CO adsorption experiments of \citet{Oncel_etal:2006}, which is indeed the case \citep{VanpouckeBrocks:2010b}.

The real issue of interest is something else: the various model assumptions, errors of omission, and idealizations harm the realism of the model and the direct, physical interpretation of the results. In the end, it does not seem possible to assess to what extent $\alpha$ (in case A) and the atomic configuration (in case B) are more than empirical parameters from a statistical model; the models are intended to have a direct realistic interpretation, but all the tricks that were necessary to arrive at them have complicated---and possibly compromised---the relation between reality and those models. As \citet[p.~197]{Fujiwara:2007} points out: ``an optical model used in ellipsometry analysis merely represents an approximated sample structure, and obtained results are not necessarily correct even when the fit is sufficiently good.'' Even if it is assumed that there is a correct model (as in case B), the applied model may still be wrong if an assumption about the experimental conditions was wrong, if some approximation was unwarranted, or if the scientist just did not come up with the right type of model. If so, it is hard to establish whether the model is just slightly off, or completely wrong.

This seems to be a very strong concession: how can physics still claim to provide a method for learning about the world, if the very models it uses just obfuscate the relation to the world? The answer is that scientific knowledge does not depend on a single experiment, performed by one group of scientists with a single technique.
Additional information from independent sources is required to elucidate the relation between the physical parameter of interest and the model parameter.

Observe that if we had computed the error on $\hat{\alpha}$, it would only have informed us of error in the fitting procedure, which is the very last step in a complex chain. There are many possible reasons why the obtained value for $\alpha$ may deviate from the actual average angle of the DNA molecules on the samples. Obviously, the results for $\alpha$ should not (yet) be treated on an equal footing with those scientific facts that have been established by multiple independent teams using different research methods and model assumptions.
Despite the uncertainties associated with the study, we now have more knowledge on the tilt angle than before the study was conducted at all. At present, $\alpha$ is our best estimation for the tilt angle of DNA molecules on diamond.
Although statisticians may frown upon the common practice of ellipsometry not to calculate error bars, their absence does promote reflection on the full uncertainty, of which model uncertainty is a more profound component as compared to fitting error.
We should view the results of cases A and B as our current best attempt at answering the research question: the obtained answer is less certain than the foundations of optics or QM itself, but more reliable than a mere educated guess.


\section{Discussion and conclusions}\label{sec:Outro}
\subsection{Models in statistics vs.\ models in material science}
The problem of model-assessment is not easily quantifiable: even though various criteria (such as the goodness-of-fit relative to a distance measure) are expressible by numbers of merit, the reason to use this particular distance measure rather than a different one is not quantifiable. Although one may take into account multiple criteria to compare models, the overall result is a partial order, not a total order. Thus, the question of putting weights on the different criteria remains open for reflection. Moreover, it is well known that the relevant criteria may be conflicting: to keep the computational cost low, a model should have a minimum of free parameters; to increase the goodness-of-fit, more parameters are better. It is clear that there is a trade-off between these two criteria, but how do we establish a good equilibrium? With respect to this question, material scientists follow a different approach as compared to statisticians. In material science, the number of parameters in a model is chosen on the basis of a physical background theory and contextual information (on the sample and the measurement conditions), rather than on a statistical information criterion.

Despite their clear differences, cases A and B do support a common conclusion. In the development of science, there is a tendency to transform qualitative decision schemes into quantitative algorithms; statistics has evolved out of more basic heuristics concerning uncertainty. However, at a given point in time, there always remain issues for which no rigorous, mathematical methods are available yet. It may be tempting to avoid discussing these issues altogether, but we argue that discussing them in---necessarily---qualitative terms is to be preferred over not addressing them at all.

\subsection{Models and realism}
Because of the close link to physical theories, material scientists tend to be realists about their models. On closer inspection, this attitude does not seem fully warranted: some parameters in the models employed by material scientists are intended to correspond directly with a physical parameter, but the accumulation of approximations in the modeling and simulation makes the correspondence less self-evident. The distinction between empirical and scientific models is not as clear-cut as one might have expected: any strict classification of models in science is bound to fail.

The realistic orientation of the material scientist has drawbacks as well as advantages. A possible drawback of the realistic orientation is that it may hinder the material scientist to revise his model (or his interpretation thereof) in view of conflicting evidence. An advantage of this disposition is its strong motivational force: it suggests that there is a correct model to be found, and that it is the goal of the scientist to find it. These observations illustrate the complex interplay between construction and discovery in the activity of scientific modeling.

\subsection{Speaking the same language}
Notice that in case A, the research question comes before the choice of the characterization method: it is chosen in function of the question. In case B, however, the research question arises as a result of the chosen method: it is a direct result of the chemical insensitivity of the STM method.

It has been observed, for instance in the context of climate modeling \citep{Guillemot:2010}, that experiments and simulations need to `speak the same language'. In climate science, experiments and simulations are done by different (groups of) scientists, who have different research cultures and different traditions regarding the way in which they present their results. They report on different quantities, making it a non-trivial task to relate their results to each other---something which is nevertheless essential in their common endeavor to describe and predict the state of the climate.

In the context of SE, the `same language' requirement is fulfilled in a very natural way. As we have seen, ellipsometry is an indirect method, which always requires an extensive amount of data analysis before the spectra reveal any interesting information. As a result, both the acquisition and the analysis of ellipsometric data is usually performed by the same person, or at least by members of the same research group. Both parts of the research result in spectra that can be plotted on one and the same graph, allowing direct comparison of the two sources of results.

In this respect, the second case is more like climate science: experimental STM is performed by one research group, whereas theoretical structure calculations are performed by another. Typically, the theoreticians perform DFT calculations and only report the energies associated with different structures. Relating these energy values to obtained STM images is a non-trivial task, since the observed structures may be metastable, which means that they need not correspond to the lowest energy structure. In our case, however, the theoreticians made their calculations speak the same language as the experiments, by calculating STM images for the theoretical structures. This enabled direct comparison between the two types of result and made it possible to answer the question regarding the chemical nature of the observed NWs and their formation process.

\subsection{Process view of science}
A new-born scientific result is soft and vulnerable to refutation by additional tests. If the new result survives further scrutiny, it becomes more solid, gradually evolving into a hard scientific fact. Most new-born scientific results never make it into scientific facts. Not only do many results have to be retracted or at least revised due to further investigation, a still larger class of results is not investigated further at all. They do not `grow up' to be facts, not because of refutation, but because of neglect by the wider research community.

Looking critically at an isolated study in material science, as we have done here for two cases, may give one the impression that these results are very likely to be wrong \citep{Ioannidis:2005}. In reality, however, the studies are not isolated, but should be regarded as fibers of a larger body of a living, scientific structure. The publication of a result (the product of science) is not the end of the story (the process of science): it will have further interactions with other parts of science, or else die. The publication of a study makes it possible for other scientists to build on its results, or to test them further. In both cases, if the results are wrong, the discrepancy between them and the actual state of the world will very likely show up, upon which the results may be refuted or adjusted in the right direction. It is true that many results do not get any further attention, such that their possible falsehood will not be detected. These results, however, should be regarded as dead ends, which will soon no longer be part of the living entity that science is.

\subsection{What error values do not tell us}
In material physics, most data are reported with an estimation of the error; the error values may be based on the standard deviation for repeatedly measured data and based on the goodness-of-fit parameters in the case of simulated data that have been fitted to experimental results. In some cases, however, it does not seem possible to compare the simulation with the experimental data in a quantitative way: this is true for the two cases we have discussed in this paper.

Scientists feel comfortable about exact numbers, and---although there are some notorious pitfalls in the interpretation of statistical data---we do think it is appropriate to strive for quantitative results. The two cases we have presented here, make use of well-defined models and use quantitative methods to arrive at a result. This result itself, however, does not provide a quantitative comparison between the experiments and the simulation. At this point, scientists may start to feel uneasy: they are used to have a number to inform them of the reliability of a result. In the absence of this information, they have to judge the value of the final result based on different criteria. We claim that, in the context of material science, error bars may provide a false sense of security. For even when the errors are given, the overall reliability of the result does not only depend on them.

Numerical error bars, if possible to compute, do help to rule-out some models that are clearly inappropriate for a given application. However, small errors are not sufficient to guarantee that the model is well-chosen: the error can be made arbitrarily small by increasing the number of parameters in the model, a phenomenon known as `overfitting'. In the absence of error values, we are forced to think about other ways of establishing the quality of a simulation method. This reflection shows that the intrinsic quality of a model, its explanatory force, its predictive powers, its computability, its applicability to a certain system, and other aspects, \emph{all} play a role in the quality of simulations in material science. These factors should be taken into consideration, even for those studies that \emph{do} report error bars.

\section*{Acknowledgement}
We are very grateful to one anonymous referee whose comments greatly helped to improve this paper.



\begin{thebibliography}{22}
\providecommand{\natexlab}[1]{#1}
\providecommand{\url}[1]{\texttt{#1}}
\expandafter\ifx\csname urlstyle\endcsname\relax
  \providecommand{\doi}[1]{doi: #1}\else
  \providecommand{\doi}{doi: \begingroup \urlstyle{rm}\Url}\fi

\bibitem[Arwin(2000)]{Arwin:2000}
Arwin, H.~(2000), \textit{Ellipsometry}, in: A.~Baszkin and W.~Norde (eds.), Physical Chemistry of Biological Interfaces, Marcel Dekker, New York, 577--607.
\bibitem[Beven and Westerberg(2011)]{BevenWesterberg:2011}
Beven, K.~and I.~Westerberg (2011), On red herrings and real herrings, accepted in \textit{Hydrological Processes}, DOI = 10.1002/hyp.7963.
\bibitem[Frigg and Hartmann(2006)]{FriggHartmann:2006}
Frigg, R.~and S.~Hartmann (2006), \textit{Scientific models}, in: S.~Sarkar and J.~Pfeifer (eds.), The philosophy of
  science: an encyclopedia, Vol~2, Routlegde, London, UK, 740--749.
\bibitem[Fujiwara(2007)]{Fujiwara:2007}
Fujiwara, H.~(2007), \textit{Spectroscopic ellipsometry: principles and applications}, Wiley, Chichester, UK.
\bibitem[Gr\"{u}ne-Yanoff(2009)]{GruneYanoff:2009}
Gr\"{u}ne-Yanoff, T.~(2009), The explanatory potential of artificial societies, \textit{Synthese}, 169, 539--555.
\bibitem[Guillemot(2010)]{Guillemot:2010}
Guillemot, H.~(2010), Connections between simulations and observation in climate computer modeling, \textit{Studies in History and Philosophy of Modern Physics}, 41, 242--252.
\bibitem[{G\"{u}rl\"{u}} et~al.(2003){G\"{u}rl\"{u}}, Adam, Zandvliet, and Poelsema]{Gurlu_etal:2003}
{G\"{u}rl\"{u}}, O., O.~A.~O.~Adam, H.~J.~W.~Zandvliet and B.~Poelsema (2003), Self-organized, one-dimensional {P}t nanowires on {G}e(001), \textit{Applied Physics Letters}, 83, 4610--4612.
\bibitem[Hughes(1997)]{Hughes:1997}
Hughes, R.~I.~G.~(1997), Models and representation, \textit{Philosophy of Science}, 64, S325--S336.
\bibitem[Ioannidis(2005)]{Ioannidis:2005}
Ioannidis, J.~P.~A.~(2005), Why most published research findings are false, \textit{PLoS Medicine}, 2, e124, DOI = 10.1371/journal.pmed.0020124.
\bibitem[\"{O}ncel et~al.(2006)\"{O}ncel, van Beek, Huijben, Poelsema, and Zandvliet]{Oncel_etal:2006}
\"{O}ncel, N., W.~J.~van Beek, J.~Huijben, B.~Poelsema and H.~J.~W.~Zandvliet (2006), Diffusing and binding of {CO} on {P}t nanowires, \textit{Surface Science}, 600, 4690--4693.
\bibitem[Marjoram  et~al.(2003)Marjoram, Molitor, Plagnol, and Tavar\'{e}]{Marjoram_etal:2003}
Marjoram, P., J.~Molitor, V.~Plagnol and S.~Tavar\'{e} (2003), Markov chain Monte Carlo without likelihoods, \textit{Proceedings of the National Academy of Sciences}, 100, 15324--15328.
\bibitem[Parker(2010)]{Parker:2010}
Parker, W.~S.~(2010), Predicting weather and climate, \textit{Studies in History and Philosophy of Modern Physics}, 41, 263--272.
\bibitem[Phan and Varenne(2010)]{PhanVarenne:2010}
Phan, D.~and F.~Varenne (2010), Agent-based models and simulations in economics and social sciences, \textit{Journal of Artificial Societies and Social Simulation}, 13, URL \url{http://jasss.soc.surrey.ac.uk/13/1/5.html}.
\bibitem[Vanpoucke(2009)]{Vanpoucke:2009}
Vanpoucke, D.~E.~P.~(2009), \textit{Ab initio study of Pt induced nanowires on Ge(001)}, PhD thesis, University of Twente, Enschede, The Netherlands.
\bibitem[Vanpoucke and Brocks(2010{\natexlab{a}})]{VanpouckeBrocks:2010a}
Vanpoucke, D.~E.~P.~and G.~Brocks (2010{\natexlab{a}}), Pt-induced nanowires on {G}e(001): a density functional theory study, \textit{Physical Review B}, 81, 085410.
\bibitem[Vanpoucke and Brocks(2010{\natexlab{b}})]{VanpouckeBrocks:2010b}
Vanpoucke, D.~E.~P.~and G.~Brocks (2010{\natexlab{b}}), {CO} adsorption on {P}t-induced {G}e nanowires, \textit{Physical Review B}, 81, 235434.
\bibitem[Wenmackers(2008)]{Wenmackers:2008}
Wenmackers, S.~(2008), \textit{Morphology, functionality and molecular conformation study of CVD diamond surfaces functionalised with organic linkers and DNA}, PhD thesis, Hasselt University, Diepenbeek, Belgium.
\bibitem[Wenmackers et~al.(2008)Wenmackers, Pop, Roodenko, Vermeeren et al.]{Wenmackers_etal:2008}
Wenmackers, S., S.~D. Pop, K.~Roodenko, V.~Vermeeren \textit{et al.} (2008), Structural and optical properties of {DNA} layers covalently attached to diamond surfaces, \textit{Langmuir}, 24, 7269--7277.
\bibitem[Wenmackers et~al.(2009)Wenmackers, Bijnens, Haenen, Michiels, and  Wagner]{Wenmackers_etal:2009}
Wenmackers, S., N.~Bijnens, K.~Haenen, L.~Michiels and P.~Wagner (2009), Diamond-based {DNA} sensors, \textit{Physica Status Solidi A}, 206, 391--408.
\bibitem[Whitehead(1920)]{Whitehead:1920}
Whitehead, A.~N.~(1920), \textit{The concept of nature}, Cambridge University Press, Cambridge, UK.

\end{thebibliography}
\end{document}